\documentclass[twocolumn]{aastex62} 
\usepackage{amsmath,amstext}
\usepackage{stackengine}
 \usepackage{mathtools}
\usepackage{graphicx}
\usepackage{txfonts}

\newcommand{\noop}[1]{}

\newcommand{\be}{\begin{eqnarray}}
\newcommand{\ee}{\end{eqnarray}}

\shorttitle{On the diversity of protoplanetary disk lifetimes}
\shortauthors{Pfalzner  et al.}
\begin{document}

\title{Deriving median disk lifetimes from disk lifetime distributions}

\author[0000-0002-5003-4714]{Susanne Pfalzner} 
\affiliation{J\"ulich Supercomputing Center, Forschungszentrum J\"ulich, 52428 J\"ulich, Germany}
\affiliation{Max-Planck-Institut f\"ur Radioastronomie, Auf dem H\"ugel 69, 53121 Bonn, Germany}


\email{s.pfalzner@fz-juelich.de}


\begin{abstract}
Observations show that individual protoplanetary disk lifetimes vary from \mbox{$<$ 1 Myr} to $\gg$ 20 Myr. The disk lifetime distribution is currently unknown. For the example of a Gaussian distribution of the disk lifetime, I suggest a simple method for deducing such a disk lifetimes distribution. The median disk lifetimes inferred with this method is also shown.
\end{abstract}


\section{Introduction}
\label{sec:intro}
Planets form from the gas and dust disks surrounding young stars. These disks develop over time, and their mass decreases with the age of the star-forming region \citep[e.g., ][]{Andrews:2020}. Eventually, internal disk dissipation processes lead to the complete dispersal of the disks \citep[e.g.][]{Williams:2011,Kunitomo:2020}. 
The median disk lifetime is the primary indicator of the time available for planet formation. The standard approach is to determine the decrease in disk fraction with age in young star clusters \citep[see, for example,][]{Haisch:2001,Hernandez:2007,Pfalzner:2014,Richert:2018,Briceno:2019}. 

Ten years ago, it was still assumed that $\approx$10 Myr is the firm upper end for disk lifetimes \citep[][]{Williams:2011}.   Recently, many protoplanetary disks older than the often-derived 1--3 Myr lifetime have been identified. Interestingly, young planets have been reported in similarly aged systems. The notion of a median disk lifetime seems limited, given the large spread in individual disk lifetimes from $<<$ 1 Myr to \mbox{$>$ 30 Myr.}  In this research note, I suggest a method of determining a disk lifetime distribution and then deducing the median disk lifetime from this distribution. This method has been mentioned in \citet[][]{Pfalzner:2022} as an alternative to the standard method of determining median disk lifetimes. 

\begin{figure*}[t]
\includegraphics[width=0.48\textwidth]{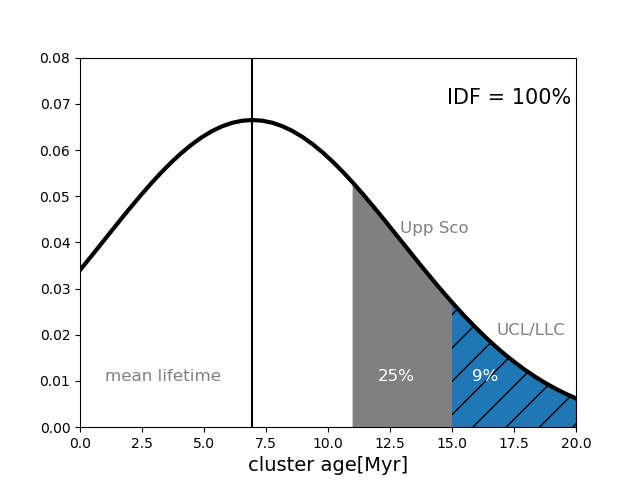}
\includegraphics[width=0.48\textwidth]{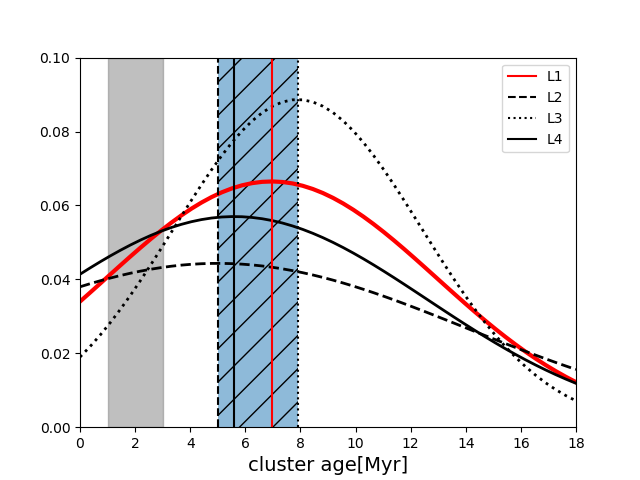}
\caption{Distribution of disk lifetimes assuming an initial disk fraction of 100\%.  The left panel  illustrates the method, the right panel shows the scatter taking the error bars into account panel. }
\label{fig:distribution_disc_lifetimes}
\end{figure*}

\section{Disk lifetime distribution}

Here I illustrate the methodology of determining a disk lifetime distribution. Currently, the disk lifetime distribution is unknown from observation. Therefore, I have to start with an assumption. If the shape of a distribution is unknown, often the first step is to assume the distribution to be  Gaussian. I follow this approach here. The shape of any Gaussian distribution can be determined if two data points are known. The methodology presented here could, in modified form, be applied to any distribution entirely determined by a limited number of points. 

For a Gaussian to be fitted, the statistical significance of these two data points has to be very high, meaning the sample size has to be large. For all the known disk fractions, the sample sizes are largest for Upper Sco (1775 stars) and UCL/LCC (3665 stars) \citep[][]{Pfalzner:2022}. Fig. \ref{fig:distribution_disc_lifetimes} left illustrates the method of determining the disk lifetime distribution for two assumed values of the disk fraction of Upper SCO of $f_d$ = 25\% with an age of $t$ = 11 Myr and the corresponding values for UCL/LCC being $f_d$ = 9\% and \mbox{$t$ = 15 Myr.} In this case, 25\% of stars have a disk lifetime longer than \mbox{11 Myr;} therefore, 25\% of the curve under the Gaussian is at values \mbox{$>$ 11 Myr.}  If one applies the equivalent constraint for UCL/LCC, the shape of the Gaussian curve is fully determined. For such a Gaussian, the maximum in the distribution is equivalent to the median disk lifetime.

\citet[][]{Luhman:2021} determined the disk fractions in Upper Sco to be 22\%$\pm$3 for the low-mass stars (M3.7-M6) in Upper Sco and  9\%$\pm$1\% for those in UCL/LLC. They assumed an age of 10 -- 12 Myr for Upper Sco and 15 -- 20 Myr for UCL/LCC. I perform a suite of fits to cover the range of possible solutions accounting for the uncertainties in these disk fraction and cluster ages. Fig.\ref{fig:distribution_disc_lifetimes} right shows the disk lifetime distribution for these values as a red line. It can be seen that the disk lifetime distribution is very broad. While the mean of the distribution is 6.94 Myr for this model, the standard deviation is with \mbox{$\sigma$= 6.5.} nearly as large. Thus the observed disks around the $\sim$ 14-Myr-old stars HD 139614 and HD 143006  \citep{Kennedy:2019,Muro:2020,Ballabio:2021} seem to be a natural consequence of such a wide distribution. 

Next, I test how sensitive the Gaussian fit is to the uncertainty of the disk fractions and cluster ages ( see Fig \ref{fig:distribution_disc_lifetimes} right panel)). Model L1 assumes the disk fractions of Upper Sco and UCL/LCC to be $f_d$(Upper SCO)= 22\% and  $f_d$(UCL/LCC)= 9\% and that these clusters are aged $t$(Upper Sco) = 11 Myr and $t$(UCL/LCC) = 17 Myr. L2 has the same parameters as L1, apart from $f_d$(Upper SCO)= 25\% and  $t$(UCL/LCC) = 15 Myr, L3 has the same parameters as L2 apart from $t$(Upper Sco) = 12 Myr and L4 has the same parameters as L1 apart from $t$(UCL/LCC) = 15 Myr. 
Importantly, each investigated set gives a mean disk lifetime significantly larger than the often assumed range of  \mbox{1 -- 3 Myr }. All models L1 -- L4) have a high standard deviation ($\sigma$ = 4.5 -- 9 Myr) (see \citet[][]{Pfalzner:2022}).

The disk lifetime distribution is very broad. 
One reason might be the age spreads within the clusters. Cluster age spreads are not very well constrained but are likely $<$ 5 Myr. Thus, they only contribute to the spread in disk lifetimes but are not the sole cause. Another reason is the disk lifetime's dependence on stellar mass \citep[][]{Pfalzner:2022}. Upper Sco shows a disk fraction of 5\%$^{+4\%}_{-3\%}$ for B7--K5.5-type stars \citep[][]{Luhman:2021} compared to 22\%$\pm$0.02 for low-mass stars (M3.7-M6). Similar, for the UCL/LLC, the disk fraction is 0.7\%$^{+0.06}_{-0.04}$ for higher-mass stars compared to 9\%$\pm$1\% for low-mass stars.

\section{Discussion}

A Gaussian distribution is often the first choice if the actual distribution is unknown. It allows for calculating the standard deviation providing more information than the mean or median values. Nevertheless, the above treatment can only be regarded as a first step. The actual distribution will be slightly skewed towards shorter lifetimes due to the mass dependency of the disk fractions. However, given the much smaller number of high-mass than low-mass stars, this is unlikely to change the distribution drastically. Similarly, external disk destruction processes can affect the disk lifetime distribution. Here again, only a small fraction of stars is affected. Nevertheless, the shape of the disk mass distribution needs further investigation in the future.

The age of Upper Sco is still debated. I also tested the case of Upper Sco being 8 Myr rather than 11 Myr old. Then the median disk lifetime reduces to 6 Myr with $\sigma$= 4.0. Thus even if Upper Sco were only 8 Myr, the median disk lifetime would be several times longer than 1 -- 3 Myr. For the high-mass stars, the median disk lifetime would only slightly change to 2.8 Myr. 

Why is the spread in disk lifetimes so large, even within the same mass bin? There are indications that the underlying reason might be connected to the internal disk substructure \citep[][]{Marel:2021}. Observations reveal that protoplanetary disks can display ring- and gap-like structures in their dust distributions. These features are associated with pressure bumps trapping dust particles at specific locations. Ring structures could prevent large (mm-sized and larger) dust from draining to the inner disk regions. In disks displaying large ring structures, disk dispersal could be delayed. This way, a fraction of high-mass stars could retain their disks considerable longer than \mbox{3 Myr} and low-mass stars, even for several tens of Myr. 

\section{Summary and Conclusion}

Disks older than 10 Myr still capable of growing planetary systems seem much more common than initially anticipated.
Thus I suggest complementing the mean/median disk lifetimes with their distributions. I derive such disk lifetime distribution based on the assumption of a Gaussian shape. These distributions lead to median disk lifetimes of 5 -- 10 Myr with ($\sigma >$ 6 Myr). My results indicate that at cluster ages $t_c$ = 0 Myr, probably some stars are diskless.

The dependence on host star mass is partly responsible for these distributions being broad. However, a wide variety of disk lifetimes is found even considering just low-mass stars. Variety in disk structure might be responsible for the observed spread \citep[][]{Michel:2021,Marel:2021}. Determining disk fractions in segregated stellar mass bins should be a prime observational aim, as it would allow the separation of stellar mass-dependent spread and other factors.

\bibliographystyle{aa} 
\bibliography{references} 

\end{document}